\documentclass[showpacs,amsmath,amssymb,aps,prb,twocolumn]{revtex4}
\usepackage{graphicx}% Include figure files
\usepackage{dcolumn}% Align table columns on decimal point
\usepackage{bm}% boldsigmasigmasigma

\begin{document}

%\draft
%\twocolumn
\title{Effect of static and dynamic disorder on electronic transport
of {\em R}Co$_{\rm 2}$ compounds: a study of
Ho(Al$_{\rm x}$Co$_{\rm 1-x}$)$_{\rm 2}$ alloys}
\author{A. T. Burkov}
\affiliation{A.F.Ioffe Physico-Technical Institute, Russian
Academy of Sciences, Sankt-Petersburg 194021, Russia.}
\author{E. Bauer, E. Gratz}
\affiliation{Institute of Experimental Physics, Vienna
Technical University, Wiedner Hauptstrasse 8-10, A-1040 Vienna,
Austria.}
\author{R. Resel}
\affiliation{Institute of Solid State Physics, Graz University of Technology,
Petergasse 16, A-8010 Graz, Austria}
\author{T. Nakama, K. Yagasaki}
\affiliation{Department of Physics, College of Science,
University of the Ryukyus, Okinawa 903-0213, Japan.}
\date{\today}

\begin{abstract}
We present experimental results on thermoelectric power ({\em
S}) and electrical resistivity ($\rho $) of pseudobinary
alloys Ho(Al$_x$Co$_{1-x}$)$_2$ ($0 \leq x \leq 0.1 $), in the temperature range
4.2~K to 300~K. The work  focuses on the effects of static
(induced by alloying) and dynamic (induced by temperature)
disorder on the magnetic state and electronic transport in a
metallic system with itinerant metamagnetic instability. Spatial
fluctuations of the local magnetic susceptibility in the alloys lead
to the development of a partially ordered magnetic ground state of the
itinerant 3{\em d}
electron system.
This results in a strong increase of the residual
resistivity and a suppression of the temperature-dependent
resistivity. Thermopower exhibits a complex temperature
variation in both the magnetically ordered and in the paramagnetic
state. This complex temperature variation is attributed to the
electronic density of states features in vicinity of Fermi
energy and to the interplay of magnetic and impurity scattering.
Our
results indicate that the magnetic enhancement of the Co 3{\em d} band
in {\em R}Co$_{\rm 2}$--based alloys upon
a substitution of Co by non-magnetic elements is mainly related to a progressive localization of the Co
-- 3{\em d} electrons caused by disorder.
We show that the magnitude of the resistivity jump at the Curie temperature
for {\em R}Co$_{\rm 2}$ compounds exhibiting a first order
phase transition is a non-monotonic function of the Curie
temperature due to a saturation of the 3{\em d}--band spin fluctuation
magnitude
 at high temperatures.

\end{abstract}
\maketitle

HoCo$_{\rm 2}$ belongs to the family of intermetallic {\em R}Co$_{\rm
2}$
cubic Laves phases ({\em R} stands for rare earth elements) which are
well known for their outstanding
magnetic properties
associated with itinerant 3{\em d} electron subsystem
\cite{Lemaire66,Levitin88,Gratz95}. Long-range
magnetic order of the 3{\em d} electron subsystem of
paramagnetic YCo$_{\rm 2}$ and LuCo$_2$ can be induced by
external magnetic field exceeding a certain critical value
$B_{\rm c}$. This critical field was found to be about 70~T for
YCo$_{\rm 2}$
\cite{Goto89} and about 77~T for LuCo$_2$ \cite{Goto90}. Partial
replacement of cobalt by aluminium leads to a decrease of
$B_{\rm c}$ and to the appearance of weak itinerant
ferromagnetism in Y(Al$_x$Co$_{1-x}$)$_2$ for $x\,>\,$0.12
\cite{Aleksandryan85,Yoshimura85}.
In the case of magnetic {\em R}Co$_{\rm 2}$ compounds, long-range
magnetic order of
the 3{\em d} electron subsystem
is induced by ordering of the localized magnetic moments of the rare
earths. A substitution of Al on Co-sites of the magnetic
{\em R}Co$_{\rm 2}$ compounds, results in a considerable increase of
the
Curie temperature \cite{Aleksandryan84}.

This behavior was
interpreted within the framework of the itinerant magnetism
by an increase of the density of states (DOS) at the Fermi
level as Co is replaced by Al. However,  the mechanism,
responsible for the increase of DOS at the
Fermi level, remains obscure. Mainly two scenarios have been
proposed: in the first, DOS increases due to
de-population of 3{\em d}-Co band when Co is replaced by Al
\cite{Aleksandryan85,Ballou93}; in the second, the DOS increase is
caused by a narrowing of the 3{\em d} band due to
an expansion of the crystal lattice in the alloys \cite{Yoshimura85,Ishiyama86}.

In the present work we use the temperature dependent thermopower
$S(T)$ to study
the mechanism of the DOS enhancement in Al-substituted {\em R}Co$_{\rm
2}$
compounds.
It has been shown that $S(T)$ high-temperature minimum,
observed in $S(T)$ of nearly all {\em R}Co$_{\rm 2}$ compounds in a
range from 150~K to 400~K, is
associated with a sharp peak in the density of states related
mainly to the Co 3{\em }d-electron density \cite{rem1}.
Particularly, it was shown
that the temperature $T_{\rm min}$, at which the thermopower has this
minimum, is a measure of the DOS peak width
\cite{Burkov88,Gratz95}.

Another unsolved problem, related to metamagnetism and charge
transport in {\em R}Co$_{\rm 2}$
compounds is the different order of the phase transition at Curie
temperature ($T_{\rm c}$) and a non-monotonic variation
of the resistivity jump ($\Delta \rho $) at $T_{\rm c}$.
Among the {\em R}Co$_{\rm 2}$ compounds, ErCo$_{\rm 2}$, HoCo$_{\rm
2}$ and DyCo$_{\rm 2}$
undergo a first order phase transition at $T_{\rm c}$, whereas it is
generally accepted that in
TbCo$_{\rm 2}$, GdCo$_{\rm 2}$, TmCo$_{\rm 2}$ and SmCo$_{\rm 2}$
the magnetic ordering is via second order transition.
The kind of the phase transition at $T_{\rm c}$ in PrCo$_{\rm 2}$ and
NdCo$_{\rm 2}$
has been a subject of recent discussion \cite{forker2003,forker2007,herrero2006,herrero2007}.
In this context the Ho(Al$_x$Co$_{1-x}$)$_2$ alloys are proper materials to study, since this alloy system includes compounds with both, first order, and
second order transition at $T_{\rm c}$.

\section{Experimental procedures}
The samples of Ho(Co$_{\rm 1-x}$Al$_{\rm x}$)$_{\rm 2}$ alloys were prepared
from pure components by melting in an induction furnace under a
protective Ar atmosphere and were subsequently annealed in
vacuum at 1100~K for about one week. The X-ray analysis showed
no traces of impurity phases.

A four--probe dc method was used for the electrical resistivity
measurements; for the thermopower measurements a differential
method was applied. Typical size of the samples was about
1$\times $ 1$\times $ 10 mm$^3$. The estimated error in the
absolute value of the electrical resistivity is $\pm $10\%. This
is mainly due to uncertainty in the sample geometry which is
closely related to the mechanical quality of the samples. The
thermopower was measured with an accuracy of $\pm $0.2~$\mu
$VK$^{-1}$.

\section{Experimental results.}
\begin{figure}[h]
\includegraphics[scale=0.9]{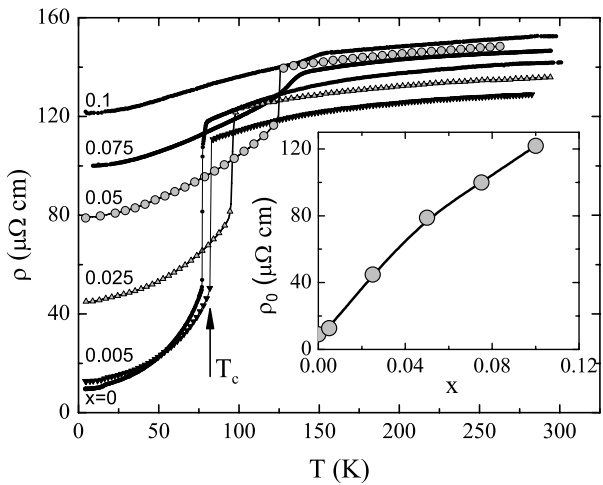}
\caption{\label{ResD} Temperature dependent resistivity of
Ho(Co$_{\rm 1-x}$Al$_{\rm x}$)$_{\rm 2}$ for various Al
concentrations. The inset shows the composition dependence of
the residual resistivity, measured at 4.2~K.}
\end{figure}

The resistivity temperature dependencies are depicted in Figure
\ref{ResD}. The residual resistivity, shown in the inset,
rapidly increases with Al content. Simultaneously, the
temperature dependent part of the resistivity decreases, so that
the high temperature total resistivity shows a comparatively
small change with composition. Room temperature values are in
the range of
130 to 150~${\rm \mu \Omega}$cm.
The Curie
temperature is marked by an abrupt change of the
resistivity of the samples with $x$=0, 0.005, 0.025, and 0.05, whereas
samples with $x$=0.075 and 0.1 clearly show a second order-like
variation of the resistivity at $T_{\rm c}$.
Data of Fig.~\ref{ResD} indicate that the boundary between first-order
and second-order transitions is between
$x$=0.05 and $x$=0.075.
 These results agree with the data of Duc et al. \cite{Duc92}.

\begin{figure}[h]
\includegraphics[scale=0.9]{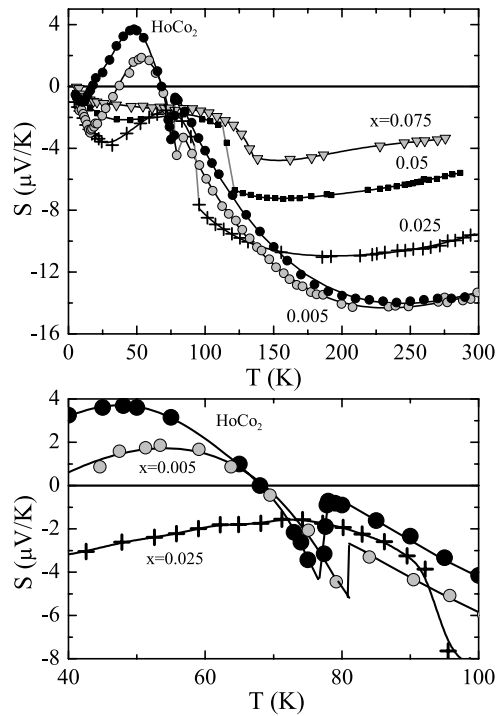}
\caption{\label{ThPw} Temperature dependent thermopower of
Ho(Co$_{\rm 1-x}$Al$_{\rm x}$)$_{\rm 2}$. The lower panel shows $S(T)$ of the
{\em x} = 0, {\em x} = 0.005 and {\em x} = 0.025 alloys in a vicinity
of Curie temperature.}
\end{figure}

Figure \ref{ThPw} presents experimental results for the
thermopower from 5~K to 300~K. Dramatic changes of $S(T)$
with the alloy composition are clearly seen in both, the paramagnetic and in the magnetically ordered state of
the alloys.
At high temperatures the $S(T)$ minimum temperature $T_{\rm min}$,
observable for pure
HoCo$_{\rm 2}$ at 250~K, and the absolute $S(T)$ values decrease with
increasing Al content.
In the low temperature region, ($T<T_{\rm c}$), $S(T)$ is also
strongly dependent on the Al-content.

At the lowest temperatures all compounds exhibit a minimum with negative
$S(T)$ values;
its temperature gradually increases with Al
concentration.
The $S(T)$ minimum, as well at the following
maximum, become suppressed for $x$=0.05 and they
almost disappear for $x$=0.075.
The Curie temperatures are associated with the discontinuous changes of
$S(T)$ for $x$=0, 0.005, 0.025, whereas
the sample with $x$=0.075 clearly shows a second order-like
variation of the thermopower at $T_{\rm c}$.

\section{Discussion.}
\subsection{Variation of the 3{\em d}-DOS upon substitution of cobalt
by non-magnetic elements.}
\begin{figure}[h]
\includegraphics[scale=0.9]{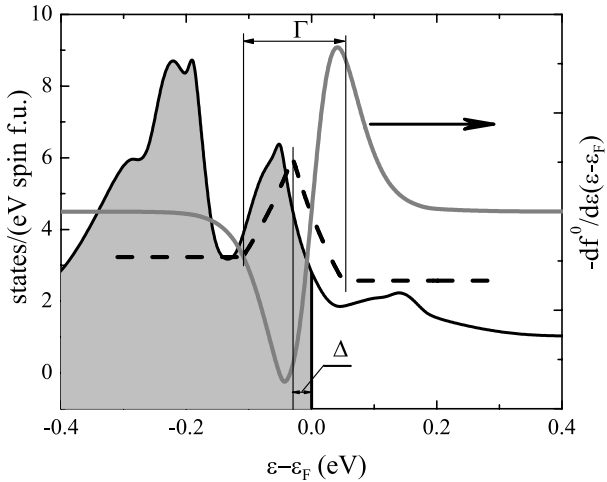}
\caption{\label{DOS} DOS of {\em R}Co$_{\rm 2}$ compounds in a
vicinity of
the Fermi level (solid line) \cite{Yamada84}. The gray line
shows $-df^0/d\varepsilon (\varepsilon -\varepsilon _F)$ at
{\em T}=300~K ($f^0$ is Fermi distribution function) to indicate the
energy region important for transport properties. The broken
line is the model state density, used for thermopower calculations.
$\Gamma$ is the width of the model DOS peak, $\Delta$ - is the
distance from Fermi level to the center of the peak.}
\end{figure}

The temperature variation of the thermopower in the paramagnetic
temperature region can be understood within the model
proposed in \cite{Burkov88,Burkov93,Gratz95}. Calculations of
the density of states, $N(\varepsilon )$ for YCo$_{\rm 2}$
\cite{Cyrot79,Yamada84,Yamada88,Tanaka98} revealed the
Fermi level, $\varepsilon _F$, lying near a sharp peak of DOS,
primarily composed of 3{\em d} states of Co (Figure \ref{DOS}).
Since Co is present in all {\em R}Co$_{\rm 2}$ compounds, it is
assumed,
that $N(\varepsilon )$ has the same general features throughout the
{\em R}Co$_{\rm 2}$ series.

It has been shown that the
high-temperature minimum, observed in $S(T)$ of nearly all
{\em R}Co$_{\rm 2}$
compounds (with the exception
of GdCo$_{\rm 2}$, DyCo$_{\rm 2}$, and SmCo$_{\rm 2}$ where 
$T_{\rm min}$ falls into
the temperature range where these compounds order magnetically),
is associated with the 3{\em d} peak of the DOS, $T_{\rm min}$ being
the measure
of the peak width~\cite{Gratz95,Burkov88}.
 
\begin{figure}[h]
\includegraphics[scale=1.1]{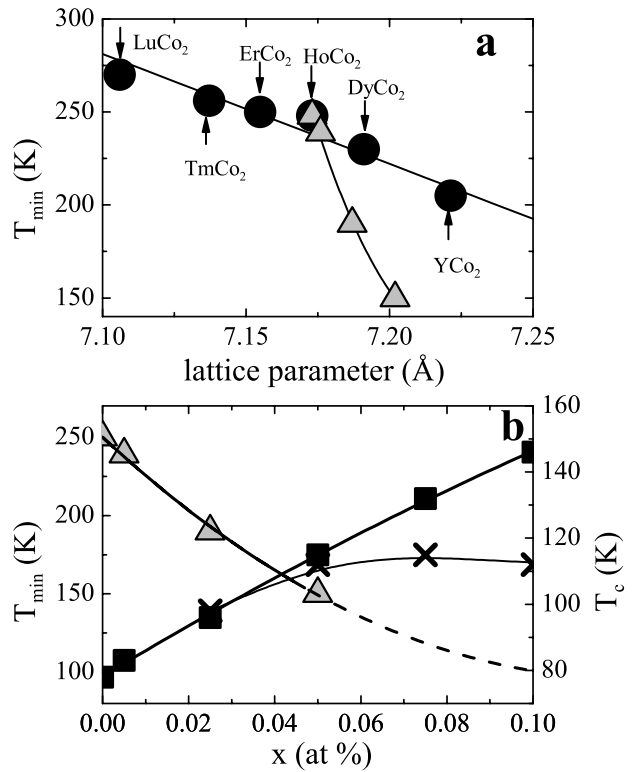}
\caption{\label{T-dcorr} {\large \bf a}. Dependence of $T_{\rm min}$
against lattice
parameter for {\em R}Co$_{\rm 2}$ compounds ({\LARGE $\bullet $} -
data
from~\cite{Gratz95}). The triangles show this work
results for  Ho(Co$_{\rm 1-x}$Al$_{\rm x}$)$_{\rm 2}$ alloys.
{\large \bf b.} $T_{\rm min}$ (triangles) and Curie
temperature $T_{\rm c}$ ({$\blacksquare $}) against composition for
Ho(Co$_{\rm 1-x}$Al$_{\rm x}$)$_{\rm 2}$. 
{\bf \large $\times $} -- $T_{\rm c}$ against composition for
Ho(Co$_{\rm 1-x}$Si$_{\rm x}$)$_{\rm 2}$ alloys according to Ref.~\cite{Duc97}.}
\end{figure}

Since the width
of an itinerant electronic band depends on the extent of the
corresponding atomic orbital overlapping, one should expect that the
3{\em d} peak width increases
when the lattice parameter decreases. The lattice parameter
within {\em R}Co$_{\rm 2}$ compounds reveals considerable variation,
being
smallest for ScCo$_{\rm 2}$ and largest for NdCo$_{\rm 2}$. This implies
that the 3{\em d} DOS peak width is largest in case of ScCo$_{\rm 2}$
and
smallest for NdCo$_{\rm 2}$ with corresponding variation of
$T_{\rm min}$.
This assumption has been confirmed by a good correlation
between $T_{\rm min}$ and the lattice constants of the 
{\em R}Co$_{\rm 2}$
compounds \cite{Gratz95}.
Figure~\ref{T-dcorr}a reproduces, in
part, this correlation.

The decrease of $T_{\rm min}$ with increasing Al content in
Ho(Co$_{\rm 1-x}$Al$_{\rm x}$)$_{\rm 2}$ alloys (Fig. \ref{T-dcorr}b)
indicates a narrowing of the 3{\em d}
DOS peak.
Partially this narrowing can be connected with the increase of the
lattice parameter in the alloys.
However, the expansion of the crystal lattice is insufficient to entirely
account for  the narrowing of the DOS peak and the observed
increase of the Curie temperature. If the decrease of
$T_{\rm min}$ was related only to the lattice expansion,
then one would expect that $T_{\rm min}$ in the alloys decreases
roughly at the
same rate as in pure {\em R}Co$_{\rm 2}$ compounds.
However, the results for
Ho(Co$_{\rm 1-x}$Al$_{\rm x}$)$_{\rm 2}$ reveal a much stronger
dependence of $T_{\rm min}$ on
the lattice parameter (Fig.\ref{T-dcorr}a), implying that some additional factor is responsible
for the narrowing of the peak width and increase of $T_{\rm c}$.

Literature results on 
Y$_{\rm 1-t}$Lu$_{\rm t}$(Co$_{\rm 1-x}$Al$_{\rm x}$)$_{\rm 2}$~\cite{Gabelko91} alloys with
invariable lattice parameter 
show, that at small {\em x} (up to about {\em
x}=0.06--0.1) the critical field for metamagnetic transition in these alloys decreases at 
the same rate as in Lu(Co$_{\rm 1-x}${\em M}$_{\rm x}$)$_{\rm 2}$ ({\em M} = Ga, Sn,
Al)~\cite{Murata91,Murata93} and
Y(Co$_{\rm 1-x}$Al$_{\rm x}$)$_{\rm 2}$~\cite{Michels90,Wada90} alloys, where lattice
parameter varies considerably. 
Similarly, in Ho(Co$_{\rm 1-x}$Si$_{\rm x}$)$_{\rm 2}$
diluted alloys~\cite{Duc97} with ivariable lattice parameter
and in Ho(Co$_{\rm 1-x}$Al$_{\rm x}$)$_{\rm 2}$ alloys  $T_{\rm c}$
increases with $x$ at the same rate, see Fig.~\ref{T-dcorr}b.
This indicates, in agreement with our results, that lattice expansion
can not be 
the main cause for the magnetic enhancement of the alloys.
On the other hand, the 3{\em d} band de-population model also can not
explain these results since 
the substituting elements have different configurations of
the outer electronic shells.

The results show that the mechanism, which causes the
decrease of $B_{\rm c}$ 
in non-magnetic compounds (or increase of $T_{\rm c}$ in
the magnetic ones) does depend at small {\em x} neither on the
type of substituting element nor on the lattice expansion.
What all above mentioned alloys have in common,
is disorder within the Co-sublattice due the
substitution of Co by other elements. We therefore conclude,
that the principal reason for the narrowing of the 3{\em d} band 
in Ho(Co$_{\rm 1-x}$Al$_{\rm x}$)$_{\rm 2}$
for low Al contents, is the
increasing localization of the 3{\em d} electrons caused by
disorder within Co-sublattice.
Experimental results on transport in Y(Co$_{\rm 1-x}$Al$_{\rm x}$)$_{\rm 2}$
alloys has led us to a similar conclusion \cite{Nakama2000}.
A different situation takes place when the impurity element has
unfilled 3{\em d}-shell, like Ni and Fe, in that case the effects of
the change of 3{\em d} electron concentration play a dominant role
\cite{Goto94}.

Of course, one can not exclude a change of the 3{\em d} electron
concentration upon the Co-substitution in Ho(Co$_{\rm 1-x}$Al$_{\rm x}$)$_{\rm 2}$.
Moreover, the decrease
of $S(T)$ values in the paramagnetic temperature
region, as will be demonstrated later,
implies such a change.
It also follows from literature
that the behavior of Y(Co$_{\rm 1-x}${\em M}$_{\rm x}$)$_{\rm 2}$
alloys at large {\em x} (above about
0.1), such as the stabilization of a ferromagnetic ground state,
depends on the kind of substituting elements,
suggesting a hybridization of {\em s--p} and 3{\em d} states.
However, our analysis  shows that the change of the 3{\em d} electron
concentration does not play major role in magnetic enhancement of
diluted alloys. 

To summarize this part, our experimental results on  transport properties of Ho(Co$_{\rm 1-x}$Al$_{\rm x}$)$_{\rm 2}$ alloys and analysis of the results, reported in the literature, 
led us to the conclusion that the principal mechanism, responsible for
 the increase of $T_{\rm c}$ in magnetic alloys,
or a decrease of $B_{\rm c}$ in paramagnetic alloys, is the narrowing
of the 3{\em d} DOS peak due to localization of Co 3{\em d} electron
states, induced by disorder in Co sublattice of the alloys.

\subsection{Electrical resistivity.}
\begin{figure}[h]
\includegraphics[scale=0.9]{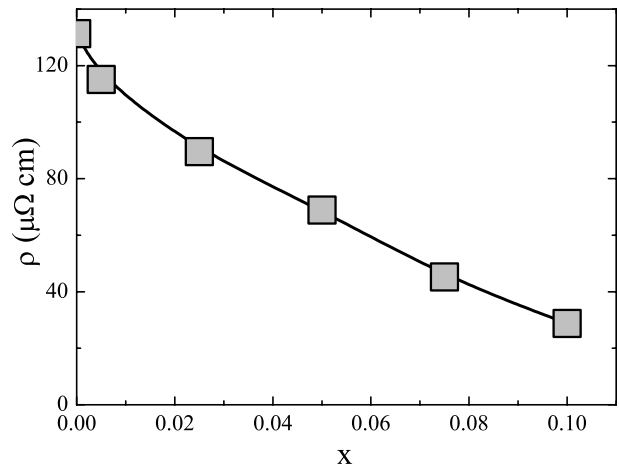}
\caption{\label{TDepR} Temperature -- induced part of resistivity ($\rho(250)-\rho(4.2)$) of 
 Ho(Co$_{\rm 1-x}$Al$_{\rm x}$)$_{\rm 2}$ alloys. }
\end{figure}

The rapid
increase of the residual resistivity and the modest variation
of high temperature resistivity with Al content are the most
characteristic features observed in the resistivity of the alloys.
The temperature
dependent part of the resistivity systematically decreases with
{\em x} as it is shown in Fig.~\ref{TDepR}. We should note that such
behavior is common for many {\em R}Co$_{\rm 2}$ - based alloys
\cite{Steiner78,Gratz87,Michels90,Duc92b,Duc95,Gratz95a,Duc97},
which include both, {\em R}(Co$_{\rm 1-x}${\em M}$_{\rm x}$)$_{\rm 2}$
and
Y$_{\rm 1-x}${\em R}$_{\rm x}$Co$_{\rm 2}$ systems.
However, there has been no commonly accepted model, able to account
for the observed variation of the temperature dependent
resistivity. The independence of the phenomenon on the constituents
of the alloys implies a common mechanism. As in the case of
the above discussed Co-localization, disorder and
metamagnetism of the magnetic Co subsystem are the natural
candidates. Baranov et al. \cite{Baranov95,Baranov2003} were
probably the first to emphasize the important role, which
disorder can play in {\em R}Co$_{\rm 2}$ - based alloys.
In a certain range of alloy composition, 3{\em d} magnetic subsystem
can be in "partially" ordered ground state, i.e. in the alloy there are
spatial regions with high and low
3{\em d} magnetization. This static magnetic disorder originates from
a combination of fluctuating exchange field (in (Y$_{\rm
1-x}${\em R}$_{\rm x}$)Co$_{\rm
2}$ alloys), or fluctuating local susceptibility (in
{\em R}(Co$_{\rm 1-x}${\em M}$_{\rm x}$)$_{\rm 2}$ alloys), and of the
metamagnetism of 3{\em d} electron
system. Scattering of conduction electrons of this static
magnetic disorder gives contribution to the alloy resistivity,
which is comparable to the high temperature limit of 3{\em d} spin
fluctuation (SF) resistivity (the resistivity due to scattering on
dynamic, i.e. temperature-induced, fluctuations of local
3{\em d} magnetization)  in
{\em R}Co$_{\rm 2}$ compounds \cite{Burkov2004}. The static magnetic
disorder can be viewed as "frozen" SF. When
temperature increases this frozen disorder melts and is replaced
by the dynamic disorder. Since the
corresponding contributions to resistivity are of similar
magnitude, the temperature variation of the total resistivity is
considerably reduced.

\begin{figure}[h]
\includegraphics[scale=0.9]{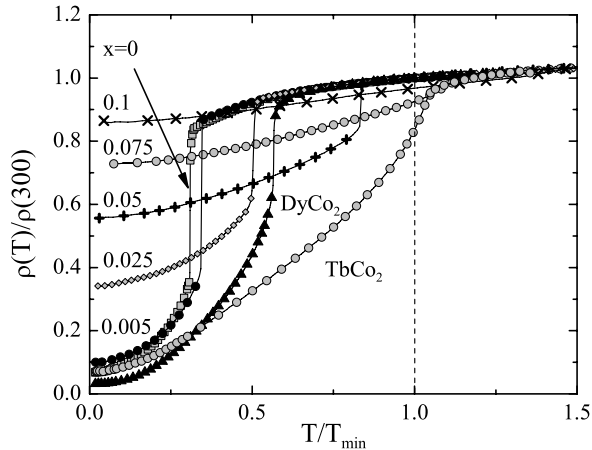}
\caption{\label{ResRel} Normalized resistivity $\rho (T)/\rho
(300)$ against normalized temperature $T/T_{\rm min}$ for
Ho(Co$_{\rm 1-x}$Al$_{\rm x}$)$_{\rm 2}$  alloys and DyCo$_{\rm 2}$, TbCo$_{\rm 2}$
compounds~\cite{rem2}.}
\end{figure}

Figure~\ref{ResRel} shows the resistivity of Ho(Co$_{\rm 1-x}$Al$_{\rm x}$)$_{\rm 2}$,
DyCo$_{\rm 2}$ and TbCo$_{\rm 2}$~\cite{rem2}, normalized to its value
at {\em T}=300~K, plotted against $T/T_{\rm min}$ to account for the
different width of the 3{\em d} band.
Since for the Ho(Co$_{\rm 1-x}$Al$_{\rm x}$)$_{\rm 2}$ with $x>$~0.075
the high--temperature minimum in $S(T)$
is not observable due to the high $T_{\rm c}$, we use an
interpolation of $T_{\rm min}(x)$,
as it is shown in Fig.~\ref{T-dcorr}b by the broken line, to obtain
$T_{\rm min}$ of these alloys.

The other two interesting features of the resistivity can be seen
from Figure~\ref{ResRel}:\\
1. The kind of the phase transition at Curie point is changed from
first to second order around $T_{\rm c}$=$T_{\rm min}$.
This is another manifestation that $T_{\rm min}$ is an
essential physical parameter, directly
reflecting DOS features. The relation between $T_{\rm min}$ and
the change of the transition order follows from the model of the metamagnetic
transition of Co subsystem \cite{Levitin88} as a consequence of the
temperature
smearing of DOS features at the Fermi
level \cite{Levitin88,Duc92}.
Essentially the same mechanism leads to the development of the
minima in $S(T)$ \cite{Burkov88}.
There is, however, an alternative theoretical model,
in which the order of a magnetic phase transition depends basically on the lattice
constant value \cite{Khmelevskyi2000}.
Since the width of the DOS peak (and, therefore, $T_{\rm min}$) is
determined in pure {\em R}Co$_{\rm 2}$ compounds
by the lattice parameter,
$T_{\rm c}/T_{\rm min} \approx 1$ represents a more general
condition for
the boundary between second and first order phase transition in these compounds.\\
2. The magnitude of the resistivity drop at the Curie temperature
$\Delta \rho (T_{\rm c})$
decreases as $T_{\rm c}$ approaches $T_{\rm min}$. This
last feature is also present in binary {\em R}Co$_{\rm 2}$ compounds,
as it can be seen comparing the resistivities of HoCo$_{\rm 2}$ and DyCo$_{\rm 2}$, 
shown in Fig.~\ref{ResRel}. The resistivity drop at
$T_{\rm c}$ of heavy {\em R}Co$_{\rm 2}$ compounds, which exhibit a
first-order
transition, has been attributed to a
sudden suppression of SF by strong molecular field
of the ordered 4{\em f} moments \cite{Gratz95}. This,
however, does not explain the decrease of $\Delta \rho (T_{\rm c})$
with increasing $T_{\rm c}$.
We propose that the origin of this decrease is the same as the origin of resistivity saturation of nearly magnetic compounds at high temperatures \cite{Ueda75,Moriya}.
In the model, put forth in \cite{Ueda75,Moriya} by Ueda and Moriya, the
SF contribution to the resistivity ($\rho _{\rm sf}$) is determined
by the {\em d} band dynamical susceptibility, which is enhanced due to
{\em d} band SF. It has been shown that the
saturation tendency, which is characteristic for the temperature
dependent resistivity of nearly magnetic materials, originates
from the strong decrease of the {\em d} band enhancement owing to
interactions among SF at high temperatures.

\begin{figure}[h]
\includegraphics[scale=1.2]{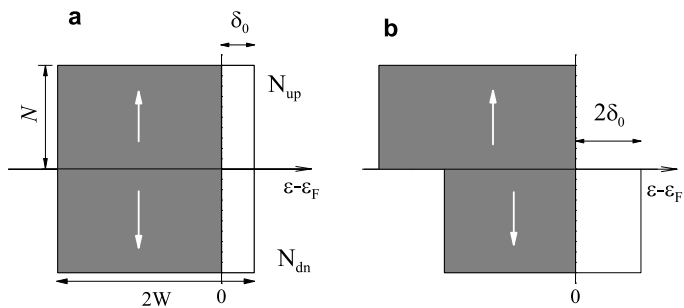}
\caption{\label{MDos}The model of 3{\em d} DOS: {\bf a.} Non polarized
state at {\em T}=0~K; {\bf b.}  Polarized state at {\em T}=0~K. 2$W$ is the 3{\em d} band width, $\delta _{\rm 0}$ is the distance from Fermi energy to the top of 3{\em d} band and $N$ -- DOS magnitude.}
\end{figure}

We use here a very simplified model with a rectangular approximation for
the Co 3{\em d} DOS (Fig.~\ref{MDos}) to illustrate the essential
physics of the resistivity
saturation and of the
decrease of $\Delta \rho (T_{\rm c})$.

We consider an enhanced 3{\em d} band paramagnetic compound (such as
YCo$_{\rm 2}$).
At {\em T}=0~K, without external field, the spin-up and spin-down
sub-bands are equally occupied,
Fig.~\ref{MDos}a, and there are no SF.
At a nonzero temperature
thermal fluctuations of the 3{\em d} sub-band occupation, amplified
by exchange interaction, result in
fluctuations of local magnetization, i.e. SF.
According to \cite{Moriya} the magnitude of the
local magnetization fluctuations
$\left( \langle m^2\rangle \right)^{\frac{1}{2}}$ increases with temperature,
however it saturates at a constant value in a high temperature limit. Both,
the saturation value and cross-over temperature, are determined by the
3{\em d} band parameters.

In magnetic {\em R}Co$_{\rm 2}$ compounds the 3{\em d} system at
{\em T}=0~K is in
high magnetization state. In the following we consider 3{\em d} band
in this case as fully polarized by the exchange field
of ordered 4{\em f} moments, as it schematically
shown in Fig.~\ref{MDos}{\bf b}, with saturation magnetic
moment $M_{\rm f} = 2N\delta _0$.
As temperature increases, there will be also
temperature-induced fluctuations of the 3{\em d} magnetization around
this polarized state.

For our qualitative model we assume that the SF resistivity is determined by the spin--spin correlation function
$$\langle m(0)m(l)\rangle - \langle m(0)\rangle \langle m(l) \rangle ,$$
where $m(0)$ and $m(l)$ are the local 3{\em d} magnetization at
lattice point with coordinates $r=0$ and $r=l$, respectively, and $l$ is
distance of order of conduction electron (s--electron) mean free path.
The scattering magnitude of conduction electron by a magnetic fluctuation is proportional to its squared magnitude, measured relative to the local background at a distance of order of conduction electron mean free path.
Therefore, the most effective will be the scattering on correlated magnetic fluctuations with largest difference of local magnetization at points $r=0$ and $r=l$.
For the paramagnetic state, this corresponds to fluctuations at $r=0$ and $r=l$  with the opposite magnetization.
While for a system in ferromagnetic ground state with fully polarized
3{\em d} band, the largest difference is
between this fully polarized state moment $M_{\rm f}$ and local fluctuation with a reduced moment.
Basing on the above arguments, we define the effective SF magnitude ($m_{\rm p}$ - paramagnetic state, $m_{\rm f}$ - ferromagnetic state), determining the SF resistivity, as:
\begin{equation}
m_{\rm p} \propto 2\left(D_{\rm up}-D_{\rm dn}\right),
\label{np}
\end{equation}
where
$$
D_{\rm up(dn)}=\int \limits _{-(2W-\delta _{\rm 0}\pm(\mp)\zeta k_{\rm B}T)}^{\infty}Nf^{\rm 0}{\rm d}\varepsilon,
$$
and
\begin{equation}
m_{\rm f}\propto M_{\rm f} - \left(D_{\rm up}-D_{\rm dn}\right),
\label{pol}
\end{equation}
with
$$
D_{\rm up}=\int \limits _{-(2W-\zeta k_{\rm B}T)}^{\infty}Nf^{\rm 0}{\rm d}\varepsilon
$$
and
$$
D_{\rm dn}=\int \limits _{-(2W-2\delta _{\rm 0}+\zeta k_{\rm B}T)}^{\infty}Nf^{\rm 0}{\rm d}\varepsilon.
$$
The definitions of $W$, $\delta _{\rm 0}$ and $N$ are given in Fig.~\ref{MDos}. $D_{\rm up}$ and $D_{\rm dn}$ are local spin-up and spin-down density, $f^{\rm 0}$ is the Fermi distribution function, and $\zeta $ is the exchange enhancement factor.

Due to the charge neutrality condition ($D_{\rm up}\! +\! D_{\rm dn}\!=\!{\rm const}$), the magnitude of the SF has an upper limit.
Both $m_{\rm p}$ and $m_{\rm f}$ have maximum value of $4N\delta _{\rm 0}$.
For the unpolarized 3{\em d} band the SF attain the maximum amplitude at a
temperature $T_{\rm m}$ satisfying condition $\zeta k_{\rm B}T_{\rm
m}=\delta _{\rm 0}$, whereas for the fully polarized 3{\em d} band the
corresponding condition is $\zeta k_{\rm B}T_{\rm m}=2\delta _{\rm
0}$.
At higher temperatures the
itinerant SF behave as localized magnetic moments
with fixed magnitude.

\begin{figure}[h]
\includegraphics[scale=0.8]{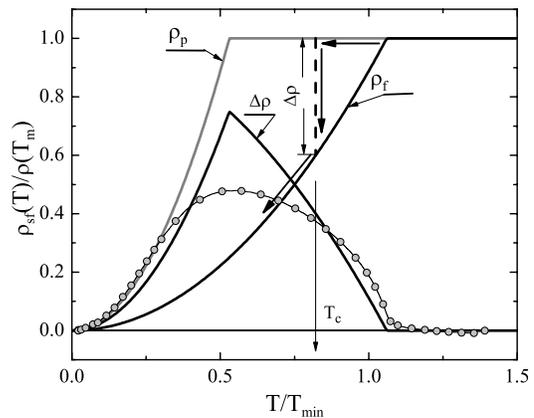}
\caption{\label{RMod}Schematic temperature dependence of
normalized $\rho _{\rm sf}$ for polarized ($\rho _{\rm f}$) and non-polarized ($\rho _{\rm p}$) states of
the 3{\em d} band and temperature dependence of resistivity jump at first order transition from non-polarized to polarized (ferromagnetic) state $\Delta \rho = \rho _{\rm
p}-\rho _{\rm f}$.
The circles
show the experimental difference between the normalized resistivities of
paramagnetic YCo$_{\rm 2}$ and ferrimagnetic
TbCo$_{\rm 2}$~\cite{rem2}. The temperature is normalized to $T_{\rm
min}$ of TbCo$_{\rm 2}$ = 214~K~\cite{Gratz95}. T$_{\rm m}$=$\frac{2\delta _{\rm 0}}{\zeta k_{\rm B}}$.}
\end{figure}

The corresponding SF resistivity follows from Eqns.~\ref{np} and \ref{pol}:
\begin{displaymath}
\rho _{\rm p} \propto \left( m_{\rm p}\right)^{\rm 2}\propto \left\{
\begin{array}{ll}
\left(4N \zeta k_{\rm B}T\right)^{\rm 2} & \textrm{if $T\leqslant \frac{\delta _{\rm 0}}{\zeta k_{\rm B}}$} \\
\left(4N\delta _{\rm 0}\right)^{\rm 2}& \textrm{if $T > \frac{\delta _{\rm 0}}{\zeta k_{\rm B}}$}
\end{array}\right.,
\end{displaymath}
for the non-polarized (paramagnetic) state, and
\begin{displaymath}
\rho _{\rm f} \propto \left( m_{\rm f}\right)^{\rm 2}\propto \left\{
\begin{array}{ll}
\left(2N \zeta k_{\rm B}T\right)^{\rm 2} & \textrm{if $T\leqslant \frac{2\delta _{\rm 0}}{\zeta k_{\rm B}}$} \\
\left(4N\delta _{\rm 0}\right)^{\rm 2}& \textrm{if $T > \frac{2\delta _{\rm 0}}{\zeta k_{\rm B}}$}
\end{array}\right.,
\end{displaymath}
for the polarized (magnetically ordered) 3{\em d} band.
Expression for $\rho _{\rm p}$ gives a roughly correct representation of the overall temperature
dependence of the SF resistivity of paramagnetic {\em R}Co$_{\rm 2}$ compounds: it increases as $T^{\rm 2}$ at low temperatures and saturates to a temperature-independent value at high temperatures \cite{Gratz95}.
Both, $\rho _{\rm p}$ and $\rho_{\rm f}$ are displayed in Fig.~\ref{RMod}.
In {\em R}Co$_{\rm 2}$ compounds with a first-order magnetic phase
transition, the 3{\em d} band undergoes at $T_{\rm c}$ a transition from the
non-polarized to the polarized state (shown schematically by the arrows in Fig.~\ref{RMod}), with corresponding change of
the SF resistivity: $\Delta \rho =\rho _{\rm p}-\rho _{\rm
f},$ which is also presented in Fig.~\ref{RMod}.
For comparison the experimental difference between the normalized resistivity of paramagnetic
YCo$_{\rm 2}$~\cite{rem2}  (a representative of a non-polarized state of the 3{\em d} band) and of ferrimagnetic
TbCo$_{\rm 2}$~\cite{rem2} (a representative of the polarized state of the 3{\em d} band) is shown on this figure too.
In spite of very schematic model, used to calculate $\Delta \rho $, there is close correspondence
between theoretical and experimental curve.

\begin{figure}[h]
\includegraphics[scale=0.9]{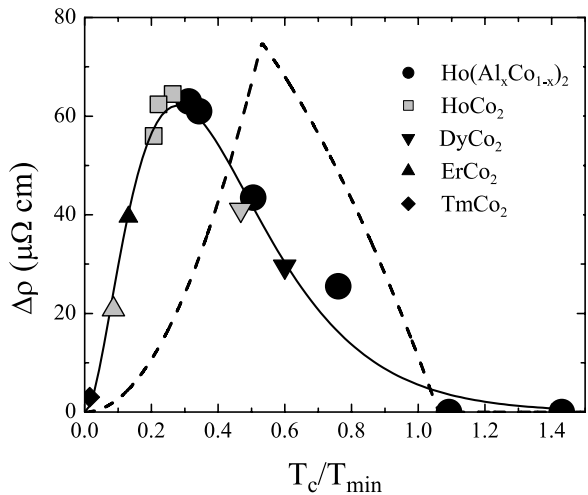} \caption{\label{DelRho}
The resistivity drop
$\Delta \rho $ at Curie temperature of Ho(Co$_{\rm 1-x}$Al$_{\rm x}$)$_{\rm 2}$
alloys and of some {\em R}Co$_{\rm 2}$ compounds~\cite{rem2,nakama99}. Broken line is
theoretical $ \Delta\rho $, scaled with $\rho (T_{\rm m})$=100~$\mu \Omega$~cm.
The gray symbols stand for data under pressure for corresponding compounds \cite{Hauser98}.}
\end{figure}

The experimental resistivity jump observed at $T_{\rm c}$ of {\em R}Co$_{\rm 2}$
and Ho(Co$_{\rm 1-x}$Al$_{\rm x}$)$_{\rm 2}$  is
presented in Fig.~\ref{DelRho}. $\Delta\rho (T_{\rm c})$ is shown
for pure TmCo$_{\rm 2}$~\cite{nakama99}, ErCo$_{\rm 2}$,
HoCo$_{\rm 2}$, DyCo$_{\rm 2}$~\cite{rem2} compounds and for Ho(Co$_{\rm 1-x}$Al$_{\rm
x}$)$_{\rm 2}$ alloys. Also shown are the results for ErCo$_{\rm
2}$, HoCo$_{\rm 2}$ and DyCo$_{\rm 2}$ under pressure up to 16~kbar \cite{Hauser98}.
The resistivity drop is plotted against
$T_{\rm c}/T_{\rm min}$ to account for the different width of Co-3{\em d} band.
We do not have experimental data on thermopower of all {\em R}Co$_{\rm
2}$ under pressure, therefore the same value of $\frac{dT_{\rm
min}}{dP}=0.7$~K/kbar was used to calculate $T_{\rm min}$ for
compounds under pressure, shown in Fig.~\ref{DelRho}.
$\frac{dT_{\rm min}}{dP}$ was obtained experimentally for a
(Gd$_{\rm 0.01}$Y$_{\rm 0.99}$)Co$_{\rm 2}$ alloy~\cite{NakamaP}. As
it can be seen from the Fig.~\ref{DelRho} there is qualitative
agreement between theoretical and experimental values of $\Delta
\rho. $  The model shows two important features in the
resistivity behavior: the saturation at high temperatures, and
the decrease of $\Delta \rho (T_{\rm c})$ 
with increasing $T_{\rm c}$. The important parameters of the
model are the bandwidth 2$W$ and the distance from the Fermi level
to band edge $\delta _0$. They can be estimated in the following
way. The empty portion of the 3{\em d} band can be evaluated from the
experimentally observed saturation Co-moment {\em M}=1~$\mu _{\rm
B}$/Co for {\em R}Co$_{\rm 2}$ compounds. From this value it follows
that $\delta _0 \approx 0.1W$. Using the experimentally observed
temperature range in which $\Delta \rho$ is non-zero
(Fig.~\ref{DelRho}) as the measure for $\delta _0$ (200~K), and
taking $\zeta =7$ \cite{Tanaka98} (which gives $\delta _0
\approx 0.12$~eV), we get for Co--3{\em d} bandwidth 2{\em W} about 2.4~eV,
in a very reasonable agreement with the value of 2.5~eV,
according to the result of Ref.~\cite{Tanaka98}.

\begin{figure}[h]
\includegraphics[scale=0.9]{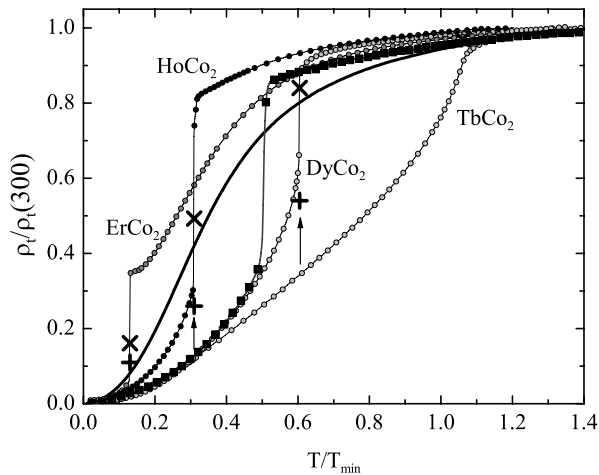} \caption{\label{RCo2-res}
Normalized resistivity of {\em R}Co$_{\rm 2}$ compounds~\cite{rem2}  and Ho(Co$_{\rm 0.975}$Al$_{\rm 0.025}$)$_{\rm 2}$
alloy ({\tiny $\blacksquare $}). The solid line is
the resistivity of YCo$_{\rm 2}$. $\rho _{\rm t}=\rho (T)-\rho (4.2)$. {\bf \Large $\times$} and {\bf $+$} indicate resistivity
at $T_{\rm c}$, including corrections for $\rho _{\rm spd}$ and $\rho _{\rm sw}$ in paramagnetic and ferrimagnetic phase, respectively.}
\end{figure}

The model DOS used here to demonstrate the mechanism of $\Delta
\rho (T_{\rm c})$ decrease with increasing $T_{\rm c}$, is obviously too
oversimplified to account for the very details of 
$\rho _{\rm sf}(T)$ and $\Delta \rho(T_{\rm c})$. The omission of 4{\em f} -- 3{\em d}
exchange interaction is less obvious. This effect, however, is clearly
visible, if one compares $\rho (T)$ of paramagnetic and magnetic
{\em R}Co$_{\rm 2}$ compounds, depicted in Fig.~\ref{RCo2-res}.
For a hypothetical magnetic {\em R}Co$_{\rm 2}$ compound, whose
resistivity includes only SF, phonon, and impurity
contributions, its normalized resistivity above $T_{\rm c}$
would coincide with the normalized resistivity of paramagnetic
YCo$_{\rm 2}$, while below $T_{\rm c}$ it should be similar to the
normalized resistivity of TbCo$_{\rm 2}$. However, as we can see
from Fig.~\ref{RCo2-res}, the normalized resistivity of the
other magnetic {\em R}Co$_{\rm 2}$ compounds deviates on approach to
$T_{\rm c}$ from the resistivity of TbCo$_{\rm 2}$ at $T<T_{\rm c}$ and
from the resistivity of YCo$_{\rm 2}$ at $T>T_{\rm c}$. In part,
this deviation refers to additional contributions owing to
scattering on 4{\em f} magnetic moments in magnetic {\em R}Co$_{\rm 2}$.  At $T>T_{\rm c}$, the
corresponding contribution, $\rho _{\rm spd}$, is independent of
temperature. Therefore, the normalized resistivity of a magnetic
compound: $$\frac{\rho (T)}{\rho (300)}=\frac{\rho _{\rm YCo_2}(T)
+ \rho _{\rm spd}}{\rho _{\rm YCo_2}(300) + \rho _{\rm spd}}$$
differs from the normalized resistivity of YCo$_{\rm 2}$.

Below $T_{\rm c}$, instead of $\rho _{\rm spd}$ there is a
temperature--dependent 4{\em f} spin-wave
contribution to the resistivity, $\rho _{\rm sw}$. It increases
with temperature, reaching at $T_{\rm c}$ a maximum value. 
In ferromagnetic metals with second-order phase transition,
$\rho _{\rm sw}(T_{\rm c})\approxeq \rho _{\rm spd}$, however in
{\em R}Co$_{\rm 2}$ with first-order transition it should be strongly
suppressed by the 3{\em d} exchange field just below the transition.
The estimated resistivity at $T_{\rm c}$, including
contributions of $\rho _{\rm spd}$ or $\rho _{\rm sw}$ is shown
in Fig.~\ref{RCo2-res}. For these estimations we use $\rho _{\rm
sw}(T_{\rm c})=\rho _{\rm spd}$, with $\rho _{\rm spd}$ from
Ref.~\cite{Gratz95}. Corrections due to $\rho _{\rm sw}$ are
certainly overestimated. Nevertheless, it is clear from
Fig.~\ref{RCo2-res} that these corrections can not account for
the observed deviations, especially above $T_{\rm c}$. The
origin of these deviations is the mutual interaction between the
4{\em f} and the 3{\em d} magnetic systems. Below $T_c$, the spin-wave
excitations in the 4{\em f} system amplify the itinerant SF of the 3{\em d}
polarized band, leading to strong increase of the resistivity on
approaching $T_{\rm c}$. In the paramagnetic state there is a
similar mutual amplification of 4{\em f} and 3{\em d} SF on approach to
$T_{\rm c}$ from higher temperatures. This amplification increases
with decreasing $T_{\rm c}$ since the susceptibility of the
localized 4{\em f} moments increases with decreasing temperature.
\subsection{Thermopower.}
$S(T)$ above $T_{\rm c}$ is related to DOS features in vicinity of
the Fermi energy. Results for Ho(Co$_{\rm 1-x}$Al$_{\rm
x}$)$_{\rm 2}$ alloys provide further support to the model
outlined in our previous publications \cite{Burkov88,Burkov93}.

We use similar approach and the model DOS, depicted in
Fig.~\ref{DOS} by the dotted line, to explore the dependence of
the thermopower on Co 3{\em d} band parameters. According to Mott's
{\em s}--{\em d} model, the conductivity is due to high mobility {\em s}--electron
states, from which the charge carriers are scattered into
3{\em d} states with low mobility and, effectively, are eliminated
from conduction process. The scattering probability of
conduction electrons is proportional to the state density value,
into which an electron is scattered. Therefore, within {\em s}--{\em d}
model, the {\em s}--electron scattering probability $P $ is expected to
depend on the electron energy as:
\begin{equation}
P (\varepsilon -\varepsilon _{\rm F}) \propto N_{\rm
d}(\varepsilon - \varepsilon _{\rm F} )+R
\label{s-d_mu}
\end{equation}
where $R$ is due to other than {\em s}--{\em d} scattering transitions. This
relation is valid for elastic or quasi-elastic scattering.

Thermopower is expressed as \cite{Burkov93,Barnard72}:
\begin{equation}
S(T)\,=\,-\frac{1}{\vert e\vert T}\, \frac{\int \limits _{0}^{\infty
}\omega(\varepsilon ,T)\left( -\frac{\partial f^0}{\partial \varepsilon}\right)
(\varepsilon -\varepsilon _{\rm F} )\, d\varepsilon}{\sigma(T)}, %\nonumber
\label{S}
\end{equation}
where
\begin{equation}
\sigma (T)\,=\,\int \limits _{0}^{\infty }\omega (\varepsilon ,T)
\, \left( -\frac{\partial f^0}{\partial \varepsilon }\right) \,
d\varepsilon %\nonumber
\label{sigma}
\end{equation}
and
\begin{equation}
\omega \left(\varepsilon ,T\right) = \frac{e^{\rm 2}}{12\pi ^{\rm 3}\hbar}vA(\varepsilon )\tau\left(\varepsilon ,T\right) , %\nonumber
\label{omega}
\end{equation}
where $v$ is the velocity of conduction electrons,
$A(\varepsilon )$ is the cross-sectional area of the Fermi
surface, perpendicular to the electrical field direction. $\tau
\propto \frac{1}{P}$ is the relaxation time. In the {\em s}--{\em d}
model approximation
\begin{equation}
\tau (\varepsilon -\varepsilon _{\rm F}) \propto \frac{1}{N_{\rm
d}(\varepsilon - \varepsilon _{\rm F} )+R}\quad .
\label{tau}
\end{equation}

\begin{figure}
\includegraphics[scale=1.5]{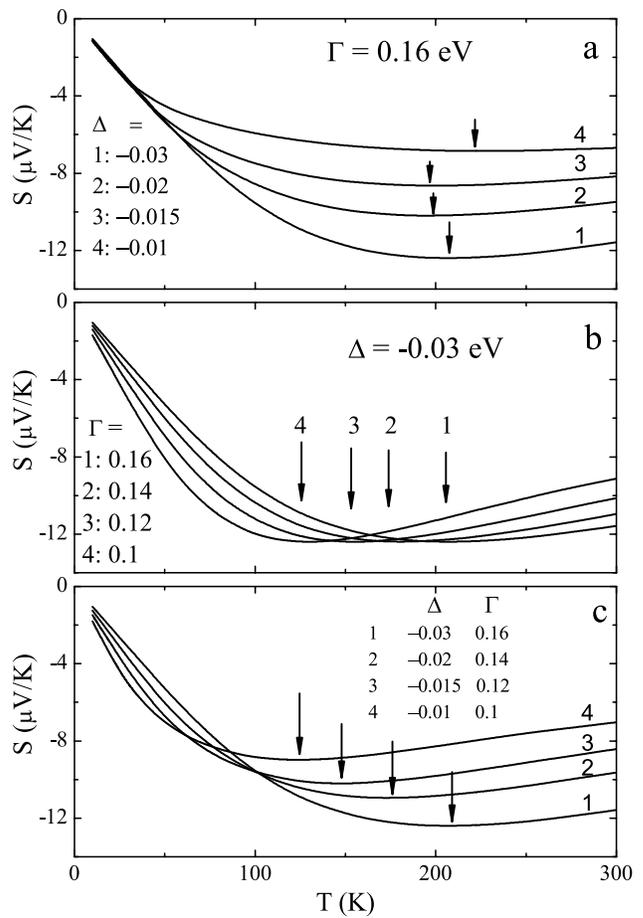}
\caption{The temperature dependencies of thermopower calculated within {\em s}--{\em d} scattering
model approximation. {\bf a.} The DOS peak width is fixed at $\Gamma $=0.16~eV,  the distance from
the Fermi energy to the center of the DOS peak, $\Delta $, varies,
simulating the de-population of the 3{\em d} band.
{\bf b.} The occupation of the 3{\em d} band $\Delta $ is fixed at -0.03~eV,
the
width of the 3{\em d} peak varies.
{\bf c.} Both, $\Gamma$ and the absolute value of $\Delta$
decrease. The arrows indicate position of the thermopower minimum.}
\label{S-Tcalc}
\end{figure}

Using equations~(\ref{S}--\ref{tau}) $S(T)$ is calculated with
the model DOS, shown in Fig.~\ref{DOS}. The results of the
calculations are shown in Fig.~\ref{S-Tcalc}. The calculations
were performed using different values of two parameters
describing the model DOS: $\Gamma$ is the width of the 3{\em d} DOS
peak, and $\Delta$ - is the distance from Fermi energy to the
center of the peak.
Two important conclusions follow from these results:\\ 1. The
position of the thermopower minimum depends mainly on the
3{\em d} peak width;\\
2. The de-population of the 3{\em d} band affects the
magnitude of the high--temperature thermopower, however has
almost no effect on the position of the thermopower minimum.

Comparison of the theoretical result of Fig.~\ref{S-Tcalc}c and
experimental data on thermopower (Fig.~\ref{ThPw}) indicates
that  both effects -- de-population and narrowing of 3{\em d} band,
play an important role determining $S(T)$ in the paramagnetic temperature range.
Both also lead to an increase of 3{\em d} DOS at Fermi level.
From the variation of parameters $\Delta $ and $\Gamma $ we can roughly estimate that
the narrowing of 3{\em d} band increases DOS at Fermi level by more than
50\%, while de-population -- by less than 20\%.
Therefore, this estimation support our conjecture, that narrowing of
3{\em d} band due to disorder plays dominant role in the magnetic
enhancement of {\em R}(Co$_{\rm 1-x}${\em M}$_{\rm x}$)$_{\rm 2}$ alloys.

The calculations of thermopower were made under assumption of elastic
scattering, which is valid at high temperatures, i.e. above
about 100~K. At lower temperatures scattering on SF is
 non--elastic. The distinction between elastic and
non-elastic scattering is essential in case of thermopower
(however it is not important for resistivity). For elastic
scattering the change of conduction electron energy satisfies
the condition: $|\varepsilon (k^{\prime}) - \varepsilon (k)| \ll
k_{\rm B}T,$ where $k^{\prime }$ and $k$ denote the final and
initial states, respectively. In non-elastic scattering the
change of the electron energy in a scattering event is of order
of $k_{\rm B}T$. On average, an electron from a state above
$\varepsilon _{\rm F}$ will be scattered into a state below
$\varepsilon _{\rm F}$ and vice versa. Therefore, while for
elastic scattering (due to negative $\frac{{\rm d}N_{\rm
d}}{{\rm d}\varepsilon}$ values at Fermi energy) $\tau
(\varepsilon
> \varepsilon _{\rm F})>\tau (\varepsilon < \varepsilon _{\rm
F})$,  for the non-elastic scattering it can be reversed: $\tau
(\varepsilon > \varepsilon _{\rm F})<\tau (\varepsilon <
\varepsilon _{\rm F})$, leading to a positive thermopower,
instead of negative values for elastic  scattering. The effect
of these different scattering regimes is clearly visible in $S(T)$
of paramagnetic {\em R}Co$_{\rm 2}$ compounds \cite{Gratz95}: with
decreasing temperature below about 100~K $S(T)$ shows a clear
tendency to change sign, which, however, is interrupted by
another process, leading to large negative thermopower values at
low temperatures.
This another process is likely caused by SF drag effect.

\begin{figure}[h]
\includegraphics[scale=0.95]{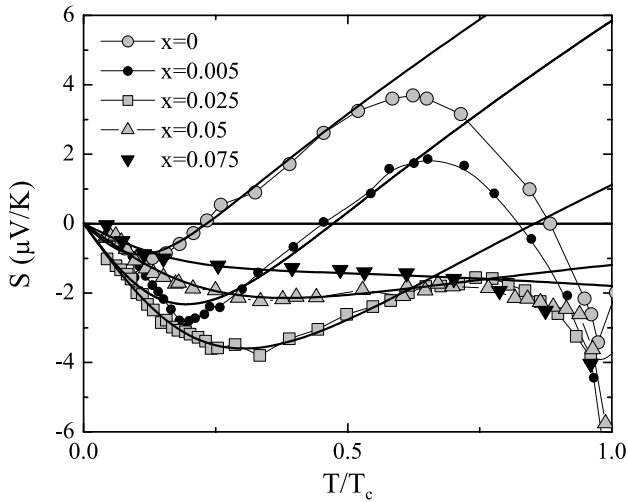}
\caption{\label{LowTr} Low temperature thermopower of
Ho(Co$_{\rm 1-x}$Al$_{\rm x}$)$_{\rm 2}$. The solid lines represent the Nordheim-Gorter expression (\ref{Stot}) with parameters,
listed in Table~\ref{tab1}.}
\end{figure}

$S(T)$ of Ho(Co$_{\rm 1-x}$Al$_{\rm x}$)$_{\rm 2}$  at
$T<T_{\rm c}$ exhibits a complicated variation as obvious
from Fig.~\ref{LowTr}.
Data imply that these temperature dependencies result from an
interplay between impurity (elastic) and spin wave (non-elastic)
scattering. For diluted alloys at low temperatures, the impurity
scattering constitutes the main contribution to thermopower.
This gives rise to negative $S(T)$ values, exhibiting an almost
linear temperature variation. At higher temperatures,
non-elastic scattering on spin waves becomes important and
results in the pronounced positive peak, especially at lower Al
content. According to the Nordheim-Gorter rule \cite{Barnard72},
thermopower of a conductor with two dominating scattering
processes, can be expressed as:
\begin{equation}
S=\frac{S_{\rm 0}\rho _{\rm 0}}{\rho } + \frac{S_{\rm sw}\rho _{\rm sw}}{\rho }, \nonumber
\end{equation}
with $\rho = \rho _{\rm 0} +\rho _{\rm sw}.$ The subscripts
refer to impurity and spin-wave scattering contributions. The
resistivity related to scattering on spin-waves follows from
$\rho _{\rm sw} = AT^2.$ We assume that both, $S_{\rm 0}$ and
$S_{\rm sw}$, are linear at low temperatures: $S_{\rm 0}=b_{\rm
0}T,$ and $S_{\rm sw}=b_{\rm sw}T.$ Based on these assumptions we
obtain for the total thermopower:
\begin{equation}
S=\frac{T}{1+\left(\frac{T}{T_{\rm 0}}\right)^2}\left[b_{\rm 0}+b_{\rm sw}\left(\frac{T}{T_{\rm 0}}\right)^2\right],
\label{Stot}
\end{equation}
where $T_{\rm 0}$ is temperature, defined from the condition:
$$AT_{\rm 0}^{\rm 2}=\rho _{\rm 0}.$$
\begin{table}[h!]
  \centering
  \caption{Fitted parameters of Eqn.~\ref{Stot}}\label{tab1}
  \begin{tabular}{|c|c|c|c|}
   \hline
    {\em x} & $b_{\rm 0}$ & $b_{\rm sw}$ & $T_{\rm 0}$ \\
\hline
    0 & -0.18 & 0.12 & 15 \\
    0.005 & -0.25 & 0.1 & 24 \\
    0.025 & -0.22 & 0.05 & 39 \\
    0.05 & -0.09 & 0.0037 & 49 \\
    0.075 & -0.055 & -0.001 & 40 \\ \hline
  \end{tabular}
\end{table}
Expression~(\ref{Stot}) was fitted to the low temperature
experimental thermopower, using $b_{\rm 0}$, $b_{\rm sw}$ and
$T_{\rm 0}$ as free parameters.  The results are shown in
Fig.~\ref{LowTr} by solid lines, and the parameter values are
listed in Table~\ref{tab1}. For diluted alloys ({\em x}=0,
{\em x}=0.005)
the parameters are reasonable and show expected dependencies on
$x$: $b_{\rm 0}$ and $T_{\rm 0}$ increase with $x$, while
$b_{\rm sw}$ is independent of $x$. At larger values of $x$,
however, the dominant low-temperature scattering mechanism is
due to static magnetic disorder, related to the formation of a
partially-ordered ground state of the 3{\em d} electron system. This
scattering is not temperature independent, which leads to a
decrease of $b_{\rm 0}$ and $b_{\rm sw}$ in concentrated alloys.

\section{Conclusion}

In conclusion, we have studied electrical resistivity and
thermoelectric power of Ho(Co$_{\rm 1-x}$Al$_{\rm x}$)$_{\rm 2}$
alloys ($0 \le x \le 0.1$) for temperatures from 4.2~K  to
300~K.
%\item
The experimental temperature dependencies of thermopower
indicate that the width of the Co 3{\em d} band decreases with Al
content. This decrease can not be accounted for only by the
increasing lattice parameter of the alloys. The analysis of
experimental results on magnetic and transport properties of
Ho(Co$_{\rm 1-x}$Al$_{\rm x}$)$_{\rm 2}$ and other {\em R}(Co$_{\rm
1-x}${\em M}$_{\rm x}$)$_{\rm 2}$ alloys led us to the conclusion that
the principal mechanism, responsible for the narrowing of the 3{\em d}
DOS peak is the
localization of the Co 3{\em d} electron states, induced by disorder
in Co sublattice of the alloys. 
The narrowing of the 3{\em d} DOS gives main contribution to enhancement of DOS at Fermi energy and to increase of $T_{\rm c}$ in magnetic alloys, or a
decrease of $B_{\rm c}$ in paramagnetic alloys.
This new mechanism of a magnetic
enhancement of the Co 3{\em d} band in {\em R}Co$_{\rm 2}$-based
alloys upon
substitution of Co by non-magnetic elements reconciles two
previous, apparently conflicting, models.

The magnitude of the resistivity jump at Curie temperature,
observable in Ho(Co$_{\rm 1-x}$Al$_{\rm x}$)$_{\rm 2}$ and
{\em R}Co$_{\rm 2}$ compounds, exhibiting first order magnetic phase
transition, is a non-monotonic function of $T_{\rm c}$. Our
simple model shows that the non-monotonous variation of the
resistivity jump with $T_{\rm c}$ is  due to the saturation of
the 3{\em d} band SF magnitude
 at high temperatures.

We postulate that fluctuations of the local magnetic
susceptibility in Ho(Co$_{\rm 1-x}$Al$_{\rm x}$)$_{\rm 2}$
 leads to development of partially ordered ground state of
the 3{\em d} magnetic subsystem, as it was shown previously for
Y$_{\rm 1-x}${\em R}$_{\rm x}$Co$_{\rm 2}$ alloys ({\em R} are
magnetic rare
earth elements). Static magnetic disorder, associated with this
partially ordered ground state, induces the huge residual
resistivities observed. With increasing temperature the static
magnetic disorder is replaced by dynamic, temperature-induced
SF. The corresponding contributions to the
resistivity have similar magnitude. Therefore, the overall
temperature variation of the resistivity in the alloys is
strongly suppressed.

Our analysis demonstrate that the temperature behavior of the
thermopower above 100~K is determined by a narrow peak in the
3{\em d} DOS at the Fermi energy. The complex behavior of the
thermopower at low temperatures results from an interplay of
elastic impurity scattering and non-elastic temperature-induced
SF scattering.

\section{Acknowledgments}
We want to thank Dr. A.Yu.Zyuzin for stimulating discussions.

This work was supported by Russian Foundation for Basic Research
under Grants 05-02-17816-a and 06-02-17047-a.


\begin{thebibliography}{99}

\bibitem{Lemaire66}
R. Lemaire,
Cobalt {\bf 32} (1966) 132; {\bf 33} 201 (1966).

\bibitem{Levitin88}
R. Z. Levitin, A. S. Markosyan,
Sov. Phys. - Usp. {\bf 31} 730 (1988).

%\bibitem{Franse93}
%J. J. M.Franse, R. J. Radwanski,
%{\em Handbook of Magnetic Materials} {\bf 7}, ed. K.H.J.Buschow (Amsterdam: North-Holland, 1993), p.307.

\bibitem{Gratz95}
E. Gratz, R. Resel, A. T. Burkov, E. Bauer, A. S. Markosyan, A. Galatanu,
J. Phys.: Condens. Matter {\bf 7} 6687 (1995).

\bibitem{Goto89}
T. Goto, K. Fukamishi, T. Sakakibara, H. Komatsu,
Solid State Commun. {\bf 72} 945 (1989).

\bibitem{Goto90}
T. Goto, T. Sakakibara, K. Murata, K. Komatsu, K. Fukamichi,
J. Magn. Magn. Mater. {\bf 90/91} 700 (1990).

\bibitem{Aleksandryan85}
V. V. Aleksandryan, A. S. Lagutin, R. Z. Levitin, A. S. Markosyan, V. V. Snegirev,
Sov.Phys.-JETP {\bf 62} 153 (1985).

\bibitem{Yoshimura85}
K. Yoshimura, Y. Nakamura,
Solid State Commun. {\bf 56} 767 (1985).

\bibitem{Aleksandryan84}
V. V. Aleksandryan, K. P. Belov, R. Z. Levitin, A. S. Markosyan, V. V. Snegirev,
JETP Letters {\bf 40} 815 (1984).

\bibitem{Ballou93}
R. Ballou, Z. M. Gamishidze, R. Lemaire, R. Z. Levitin, A. S. Markosyan,
V. V. Snegirev, Sov. Phys. JETP {\bf 75} 1041 (1993).

\bibitem{Ishiyama86}
K. Ishiyama, K. Endo,
J. Phys. Soc. Japan {\bf 55} 2535 (1986).

\bibitem{rem1} This DOS feature is responsible for the itinerant
metamagnetism of {\em R}Co$_{\rm 2}$ compounds.

\bibitem{Burkov88}
A. T. Burkov, M. V. Vedernikov, E. Gratz, Solid State Commun. {\bf 67} 1109 (1988).

\bibitem{forker2003}
M. Forker, S. M\"uller, P. de la Presa, A. F. Pasquevich,
 Phys. Rev. B {\bf 68} 014409 (2003).

\bibitem{forker2007}
M. Forker, S. M\"uller, P. de la Presa, A. F. Pasquevich, Phys. Rev. B {\bf 75} 187401 (2007).

\bibitem{herrero2006}
J. Herrero-Albillos, F. Bartolom\`e, L. M. Garc\'ia, F. Casanova, A. Labarta, X. Batlle, Phys. Rev. B {\bf 73} 134410 (2006).

\bibitem{herrero2007}
J. Herrero-Albillos, F. Bartolom\`e, L. M. Garc\'ia, F. Casanova, A. Labarta, X. Batlle, Phys. Rev. B {\bf 75} 187402 (2007).

\bibitem{Duc92}
N. H. Duc, T. D. Hien, R. Z. Levitin, A. S. Markosyan,
P. E. Brommer, J. J. M. Franse, Physica B {\bf 176} 232 (1992).

\bibitem{Burkov93}
A. T. Burkov, E. Gratz, E. Bauer, R. Resel, J. Alloys Compd. {\bf 198} 117 (1993).

\bibitem{Cyrot79}
M. Cyrot, M. Lavagna, J.Physique {\bf 40} 763 (1979).

\bibitem{Yamada84}
H. Yamada, J. Inoue, K. Terao, S. Kanda, M. Shimizu, J. Phys. F: Metal Physics {\bf 14} 1943 (1984).

\bibitem{Yamada88}
H. Yamada, Physica B {\bf 149} 390 (1988).

\bibitem{Tanaka98}
S. Tanaka and H. Harima, J. Phys. Soc. Japan {\bf 67} 2594 (1998).

\bibitem{Duc97}
N. H. Duc, T. K. Oanh, J. Phys.: Condens. Matter {\bf 9} 1585 (1997).

\bibitem{Murata91}
K. Murata, K. Fukamichi, H. Komatsu, T. Sakakibara, T. Goto,
J. Phys.: Condens. Matter {\bf 3} 2515 (1991).

\bibitem{Murata93}
K. Murata, K. Fukamichi, T. Sakakibara, T. Goto, K. Suzuki, J. Phys.: Condens. Matter {\bf 5} 1525 (1993).

\bibitem{Michels90}
D. Michels, J. Timlin, T. Mihhlisin, J. Appl. Phys. {\bf 67} 5289 (1990).

\bibitem{Wada90}
H. Wada, M. Hada, K. N. Ishihara, M. Shiga, Y. Nakamura, J. Phys. Soc. Japan {\bf 59} 2956 (1990).

\bibitem{Gabelko91}
I. L. Gabelko, R. Z. Levitin, A. S. Markosyan, V. I. Silantiev, V. V. Snegirev,
J. Magn. Magn. Mater. {\bf 94} 287 (1991).

\bibitem{Nakama2000}
T. Nakama, K. Shintani, M. Hedo, H. Niki, A. T.
Burkov, K. Yagasaki,  Physica B{\bf 281\&282} 699 (2000).

\bibitem{Goto94}
T. Goto, H. Aruga Katori, T. Mitamura, K. Fukamichi, K. Murata,
J. Appl. Phys. {\bf 76} 6682 (1994).

%\bibitem{Aoki89}
%M. Aoki, H. Yamada,
%J.Magn. Magn. Mater. {\bf 78} 377 (1989).

\bibitem{Steiner78}
W. Steiner, E. Gratz, H. Ortbauer, H. W. Gamen,
J.Phys. F: Metal Physics {\bf 8} 1525 (1978).

\bibitem{Gratz87}
E. Gratz, N. Pillmayr, E. Bauer, , G. Hilscher,
J. Magn. Magn. Mater. {\bf 70} 159 (1987).

\bibitem{Duc92b}
N. H. Duc, V. Sechovski, D. T. Hung, N. H. K. Ngan,
Physica B {\bf 179} 111 (1992).

\bibitem{Duc95}
N. H. Duc, P. E. Brommer, X. Li, F. R. de Boer, J. J. M. Franse,
Physica B {\bf 212} 83 (1995).

\bibitem{Gratz95a}
E. Gratz, R. Hauser, A. Lindbaum, M. Maikis, R. Resel, G. Schaudy, R. Z. Levitin, A. S. Markosyan, I. S. Dubenko, A. Yu. Sokolov, S. W. Zochowski,
J.Phys.: Condens. Matter {\bf 7} 597 (1995).

\bibitem{Baranov95}
N. V. Baranov, A. N. Pirogov,
J. Alloys Compd. {\bf 217} 31 (1995).

\bibitem{Baranov2003}
N. V. Baranov, A. A. Yermakov, A. Podlesnyak, J. Phys.: Condens. Matter {\bf 15} 5371 (2003).

\bibitem{Burkov2004}
A. T. Burkov, A. Yu. Zyuzin, T. Nakama, K. Yagasaki, Phys. Rev. B {\bf 69} 144409 (2004).

\bibitem{rem2} We measured resistivity of pure {\em R}Co$_{\rm 2}$
compounds (except TmCo$_{\rm 2}$) using new samples, prepared
according to the same procedure as the samples of Ho(Al$_{\rm
x}$Co$_{\rm 1-x}$) alloys. These new results agree within experimental
uncertainty with earlier data from Refs.~\cite{Gratz95,Hauser98}. 

\bibitem{Khmelevskyi2000}
S. Khmelevskyi, P. Mohn, J. Phys.: Condens. Matter {\bf 12} 9453 (2000).

\bibitem{Ueda75}
K. Ueda, T. Moriya,
J. Phys. Soc. Japan {\bf 39} 605 (1975).

\bibitem{Moriya}
T. Moriya, {\em Spin Fluctuations in Itinerant Electron
Magnetism.} Springer Series in Solid State Sciences {\bf 56}, Ed.
M. Cardona, P. Fulde, H.-J. Quisser. Springer-Verlag, Berlin, Heidelberg,
1985.

\bibitem{Hauser98}
R. Hauser, E. Bauer, E. Gratz,
Phys. Rev. B {\bf 57} 2904 (1998).

\bibitem{nakama99}  T. Nakama, K. Shintani, K. Yagasaki, A. T. Burkov, Y.
Uwatoko, Phys. Rev. B {\bf 60}, 511 (1999).

\bibitem{NakamaP}
T. Nakama, et al., unpublished.

\bibitem{Barnard72}
R. D. Barnard,
{\em Thermoelectricity in metals and alloys},
London: Tailor \& Frances, 1972.
\end{thebibliography}
\end{document}